# Intelligent data analysis based on the complex network theory methods: a case study

Olesya Mryglod, *Lviv Polytechnic National University*
(**01.12.2005**, Prof. L. Sikora, *Lviv Polytechnic National University),*
Prof. Yu. Holovatch, *Institute for Condensed Matter Physics of the National Academy of Sciences of Ukraine*

**Abstract**

The development of modern information technologies permits to collect and to analyze huge amounts of statistical data in different spheres of life. The main problem is not to only to collect but to process all relevant information. The purpose of our work is to show the example of intelligent data analysis in such complex and non-formalized field as science. Using the statistical data about scientific periodical it is possible to perform its comprehensive analysis and to solve different practical problems. The combination of various approaches including the statistical analysis, methods of the complex network theory and different techniques that can be used for the concept mapping (e.g., see [1]) permits to perform an intelligent data analysis in order to obtain underlying patterns and hidden connections. Results of such analysis can be used for particular practical problems like information retrieval within journal.

## 1. Introduction

An actual problem of modern information technologies is not only to collect but also to interpret all relevant information in different fields of life. Therefore, objects of scientific interest are the data mining methods and various approaches of intelligent data analysis. The knowledge, extracted from raw data, can be effectively used to support the decision making process in the complex systems especially where people take part in. Science is one of the examples of such complex systems. An important scientometrical problem is to evaluate and to analyze the particular scientific journals. Our goals here are the following: (i) to get the least dataset necessary and sufficient to characterize the journal, (ii) to perform the intelligent data analysis in order to obtain underlying patterns and hidden connections, (iii) to combine different methods of data analysis including the statistical analysis, the methods of the complex network theory and different techniques that can be used for the concept mapping (e.g., see [1]). The particular purpose is to apply the methods of intelligent data analysis to solve the concrete practical problem, e.g. information search within the journal. We have selected the 15-years old international scientific physical journal "Condensed Matter Physics" [1] (CMP) publishes in Ukraine. Since 2005 it has been included into the ISI master journal list and recently it has obtained the impact factor.

## 2. The analyzed data and tools

On the first step we collected all available data about (i) the authors of CMP and their affiliation, (ii) the content of the papers published in CMP, as well as (iii) the data regarding papers cited in CMP (from the reference lists) [2].

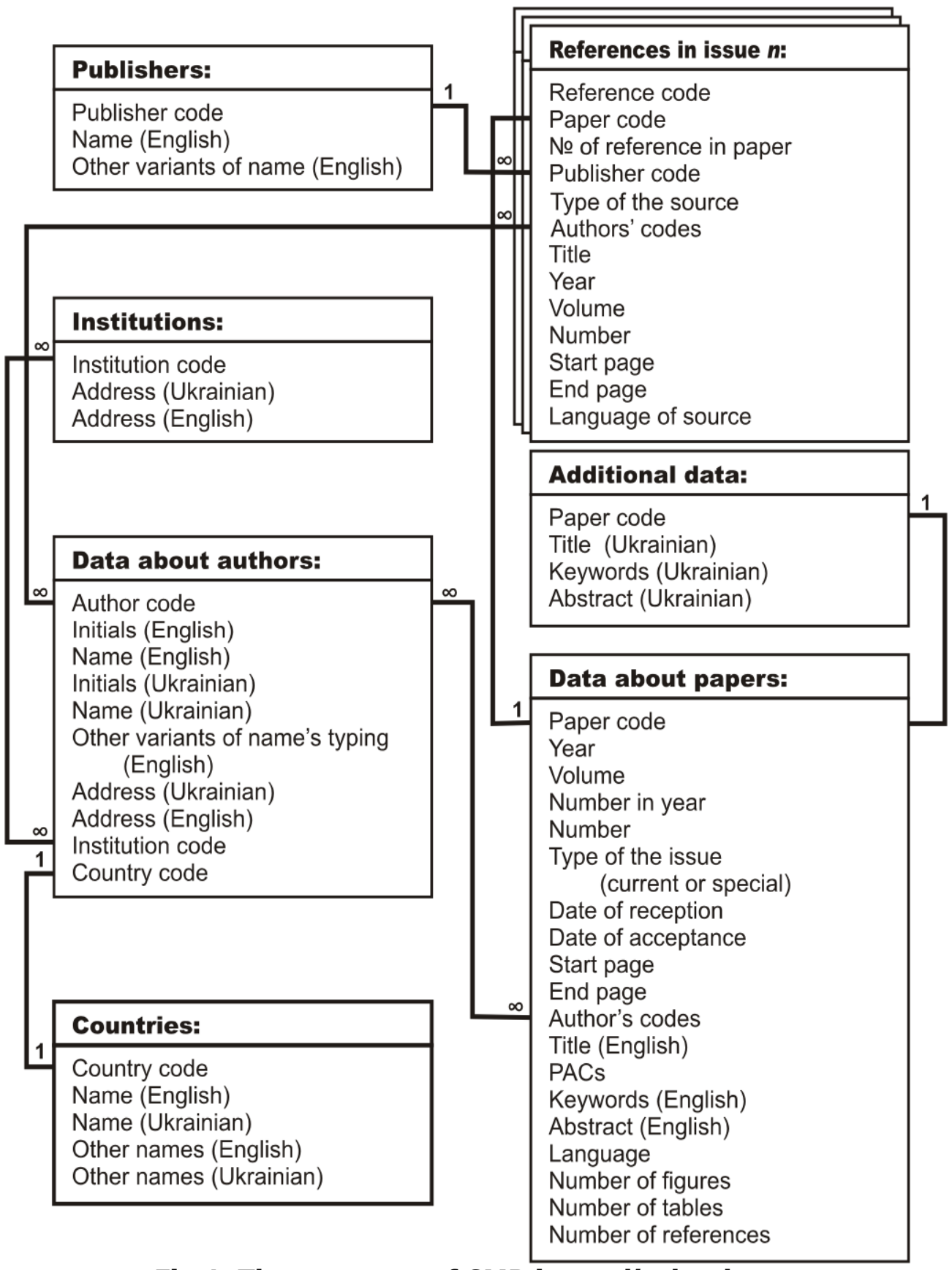

**Fig.1. The structure of CMP journal's database.**

[1] The official web-page: http://www.icmp.lviv.ua/journal

The structure of the database is shown in Fig. 1. The instruments of relational database management system Microsoft Access were used for database embedding. The Borland Delphi 7 toolkit was used provide the working with database and necessary data processing.

## 3. Application of the complex networks theory

To analyze a huge amount of interconnected data it is convenient to present these data in the form of complex network: nodes connected by edges (undirected links) or arcs (directed links) [3]. Scientometry may serve as a good example of successful application of the complex network theory. In particular, an analysis of a scientific journal can involve complex networks of several kinds: co-authorship, citation, co-citation, network of bibliographic couples. There, the interconnections between separate papers, authors, author's affiliations, etc., can be considered. Different kind of connections can link the nodes of one sort. For example, papers of the journal can be united by the authors, common citations or the same PACS numbers[2] (Fig. 2a).

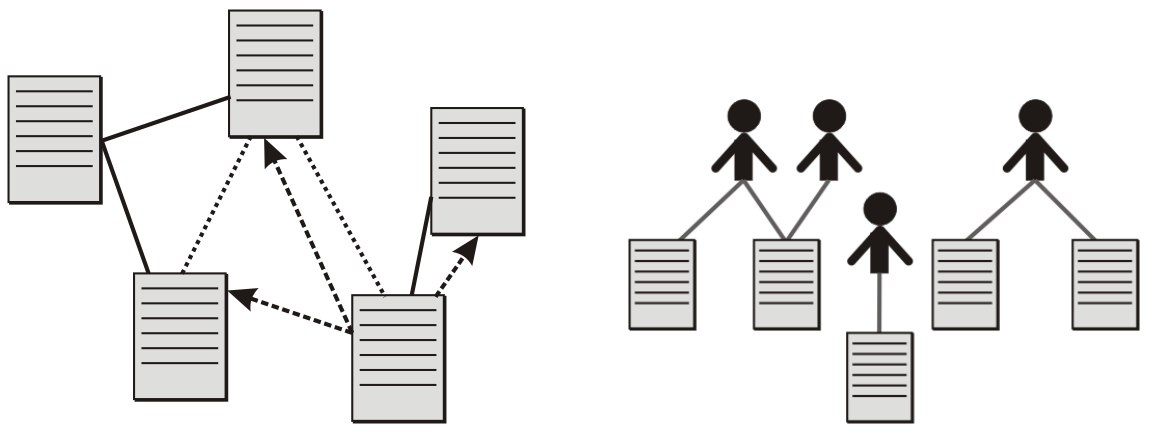

**Fig.2. Left: Schematic network of journal papers that can be connected in the following cases: they have at least one common author (solid line); one paper cites the other (dashed lines with arrows); they use the same PACS number (dotted lines). Right: Schematic example of bipartite co-authorship network that contains nodes of two sorts: authors and papers.**

Therefore, the most interesting task is to compare networks with the same nodes but various links. Moreover, the so-called bipartite networks contain nodes of two kinds, e. g. connecting together authors and their papers (Fig. 2b).

Very often only one type of network is used for evaluation purposes, the other being discarded for no obvious reasons. Our goal is to offer a comprehensive analysis of the journal based on different possible complex network interpretation of its data.

So, our next step was composition of networks based on the created database. Each network was presented as the set of nodes with their nearest neighbours (NN) and some addition information. For example, the co-authorship network was represented as in Tab. 1. The same principle was used for the rest of networks constructed. All bipartite networks were presented by their one-mode projections in a way as described below.

**Tab.1.**

**The example of the co-authorship network representation.**

| Author's ID | Nearest neighbours ID | Number of NN (degree) | number of author's papers in CMP |
|---|---|---|---|
| 100 | 10446\|4385\|4368 | 4 | 3 |
| 3672 | 3671\|3673\|3674 | 3 | 1 |

The typical complex networks analysis starts with calculation of standard network parameters like general number of nodes and links, the mean and maximal values of node degree and clustering coefficient, the mean shortest path between any two nodes and the network diameter [3]. Besides the mean and the maximum values of parameters the distribution of node characteristics for whole network are usually considered. Depending on available data it is also possible to study the time evolution of some network characteristics.

In this work the performed network analysis does not present in detail. Here, we do not present the details of an analysis (e.g. see [2]) rather concentrating on different effects based on the networks connectedness.

## 4. The network analysis

The data connected with a journal can be presented in a form of different networks depending on investigation task. We have analyzed the data on two levels considering relations between papers and authors denoted by nodes of the networks. We have studied the various kinds of interconnections between the nodes based on the co-authorship data, citations, reference and PACS numbers used in the papers, etc. Practically every kind of relationship can be presented in a form of bipartite network. Then it is convenient to separate the bipartite network and create two simple projections in a way shown in Fig. 3.

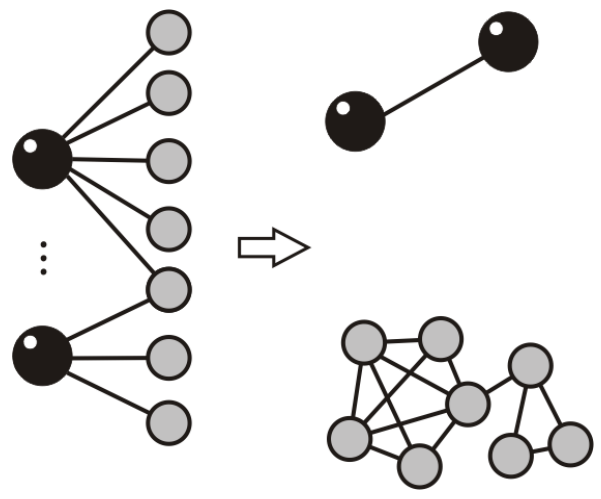

**Fig.3. The principle of separation of bipartite network to simple projections: the nodes of one sort are connected if they have at least one common node of the other sort.**

We have considered a set of networks on the level of papers and on the level of authors based on the data about CMP journal. The analysis of every built network starts with calculation of basic parameters and characteristics mentioned in the previous section. Further we describe the analyzing process of the **network of the co-authored papers as an example.**

[2] PACS is acronym for Physics and Astronomy Classification Scheme. PACS numbers, developed by the American Institute of Physics, have been used since 1975 to identify fields and sub-fields of physics.

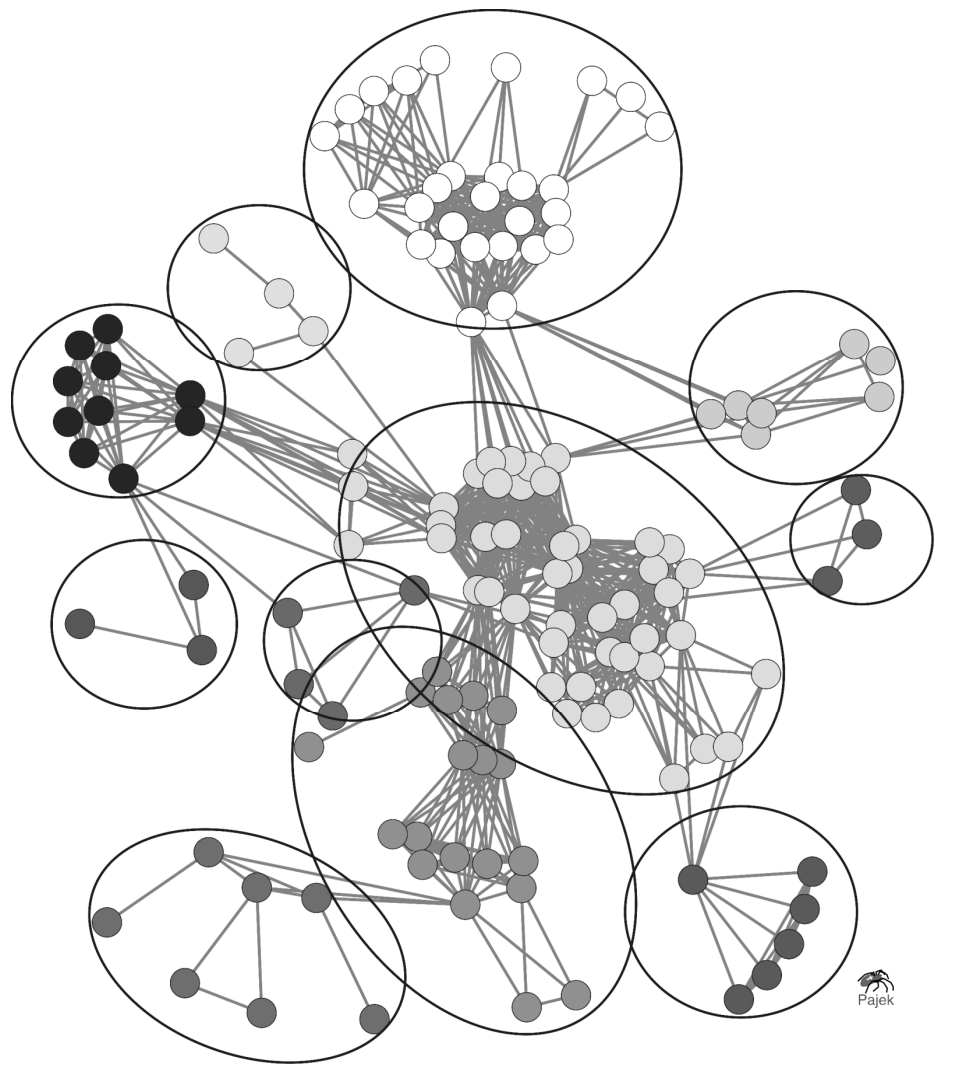

**Fig.4. The community structure of main cluster of the network of co-authored papers in the CMP journal[3].**

In this network the papers published in CMP journal during 10 years (1998–2007) are denoted by nodes and are coupled by link if they have at least one common author. The node degree in this network can be interpreted as the activity level of the paper authors within CMP journal. The connection between the papers in the terms of co-authorship in most of cases also means their relatedness in respect to their scientific directions. Therefore it is important to investigate the connected patterns within given and other networks. The largest connected cluster of the co-authored papers network contains 137 nodes (~24%). Using the Girvan-Newman algorithm [4] we have found the community structure of this core of CMP papers represented by 11 groups of nodes (Fig. 4).

We can assume that every community may represent the group of papers with similar subjects. The scientific direction of a paper can be detected using different linguistic methods based on the content analysis, but this is not the topic of our studying. The scientific direction of each paper can be formally recognized by lists of used keywords and PACS number. The ratings of used PACS numbers can be obtained for all separate clusters of the network and for all communities inside large clusters.

The other kinds of connections between CMP papers we have concerned based on the common usage of PACS numbers and citations. Depending on the task it is convenient to use different networks jointly to obtain some useful information. For example, for the subject search among the CMP papers it is helpful to know the closely related papers to specific one. Lets take for example CMP paper with ID v4n4p14 (vol. 4, No. 4, page 14 in CMP). If we are interested in it we would like to find all the related papers within CMP. Using three different kinds of networks (the network of coauthored papers, the network of papers with common PACS number and the bibliographic coupling network) we can find all neighbours of a given paper . Moreover, we can specify the so-called depth level of the neighbourhood (in our example we account all neihbours of the 1st level).

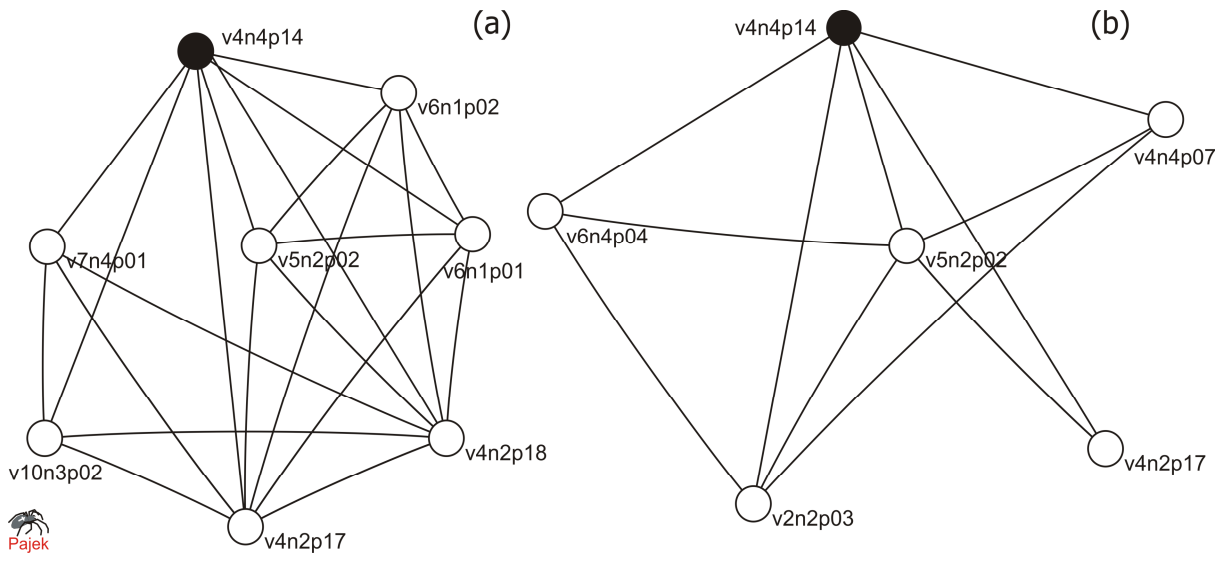

**Fig.5. The fragments of the co-authored CMP papers network (a) and the network of CMP papers with common PACS numbers (b) that contains only the paper v4n4p14 and its neighbours of specified 1st level.**

In Fig. 5a the fragment of the co-authored papers network is shown. Besides the links directly from v4n4p14 paper we can also trace the relations between its neighbours. The neighbourhood of v4n4p14 paper in the rest networks can be found in the same way (Figs. 5b,6a). Moreover, if the fragment has complex internal structure it is possible to obtain it using the Girvan-Newman algorithm [4]. The possibility to find subgroups in the network can be confirmed (or not) by the algorithm.

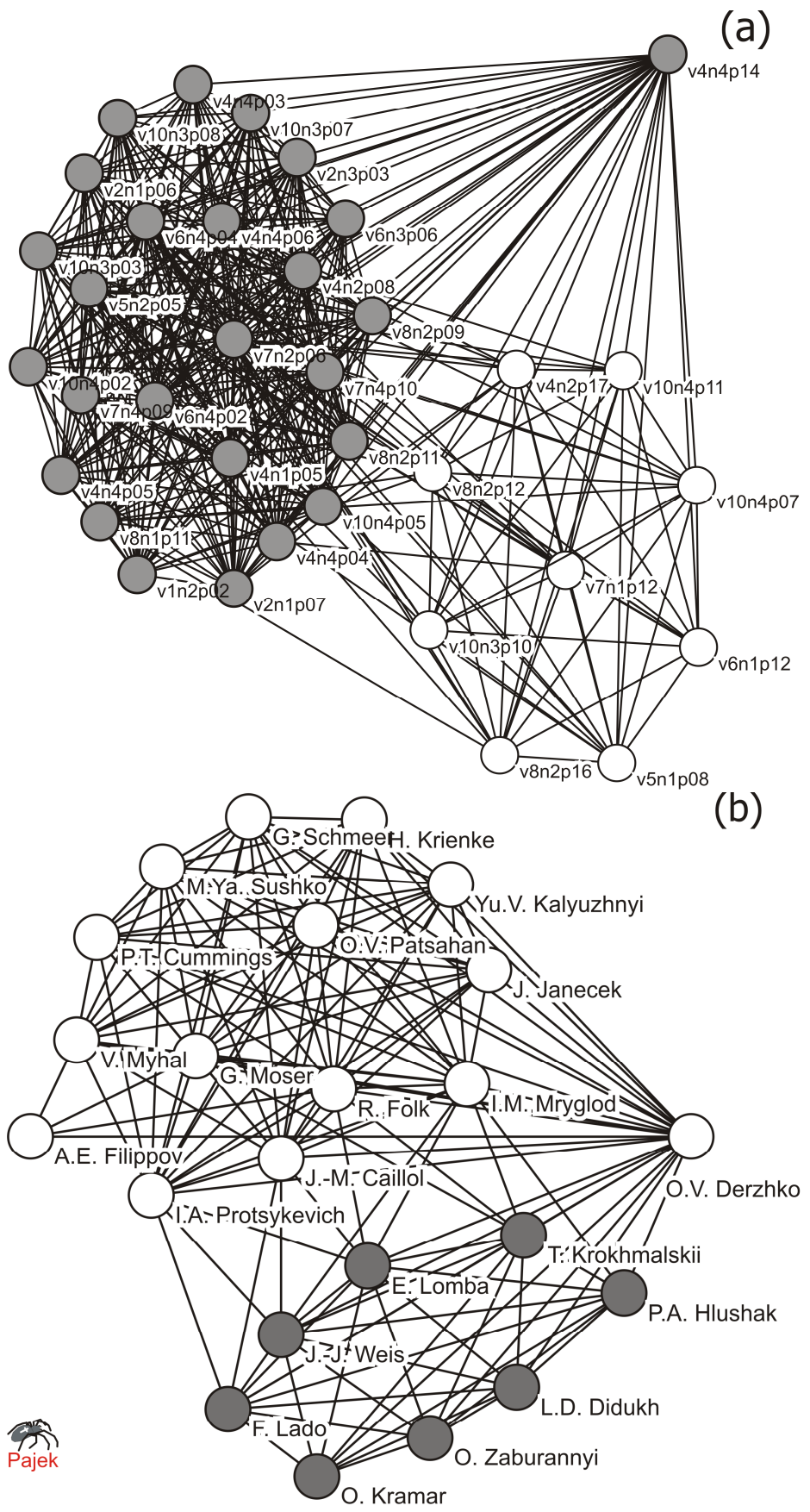

**Fig.6. The 1st level neighbourhood of paper v4n4p14 (a) and of the author 101 („O.V. Derzhko") in the networks where links mean usage of common PACS numbers. The detected communities of nodes were obtained by Girvan-Newman algorithm [4].**

[3] The network visualizations are performed by the Pajek and NetDraw softwear [4].

Two communities of the CMP papers what sharing PACS number with paper v4n4p14 are represented in Fig. 6a by different colors. The existence of the separable communities can mean the interdisciplinary character of a given paper or the usage of different methodologies. Additionally we can compare obtained lists of related papers to find common elements. In our case the paper with ID v4n2p17 is connected with v4n4p14 paper by all three kinds of relations.

Similar to the previous example, we can find the author's neighbourhood of specified depth level in different networks. For example, the neighbourhood of 1st level for authors with ID 101 (O.V. Derzhko) in the network of authors who use common PACS number in CMP papers is presented in Fig. 6b.

It is important to note that different kinds of relatedness are reflected in different networks. On the one hand in the co-authorship network the connections between the authors are provoked by themselves and also mean social relationships. In other words these connections show how author position himself in his scientific area [6]. On the other hand the closeness in the co-citation, the bibliographic coupling and the common-PACS networks is self-organized and demonstrate the author's position from the point of view of his colleges.

## 5. Conclusions

The peculiarity of our work is the application of different methods of data analysis to the local database as well as the practical use of obtained results of comparable analysis using produced and freeware software [5]. Although the limited dataset prevents from fundamental conclusions about the characteristics of built complex networks, the practically useful knowledge can be obtained. Nevertheless the analysis of CMP networks show the properties similar to analogous networks based on the much larger datasets, e.g. the scale-free nature or great level of connectivity for co-citation and co-authorship network. We show the importance of comparative analysis in order to describe the scientific periodical or to perform the information retrieval within.

### Authors:

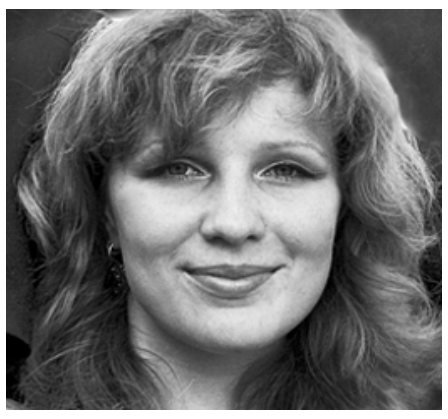

Olesya Mryglod
Lviv Polytechnic National University,
12 Bandery Str.,
79013 Lviv, Ukraine
tel. +38(097)9077557

email: *olesya.m@gmail.com.ua*

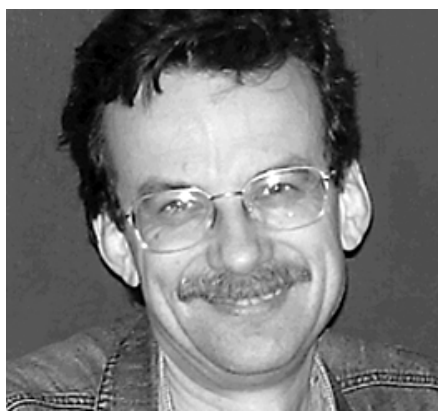

Yurij Holovatch
Institute for Condensed Matter Physics of the National Academy of Sciences of Ukraine,
1 Svientsitskii Str.,
79011 Lviv, Ukraine